\documentclass[preprint, longbibliography]{revtex4-1}

\usepackage[utf8]{inputenc}
\usepackage{amsmath, amssymb,graphicx}
\usepackage[cdot,mediumqspace,squaren]{SIunits}

\newcommand{\Ham}{\hat{H}} % Hamiltonien
\renewcommand{\vec}{\mathbf} % Vecteur
\newcommand{\e}[1]{\text{e}^{#1}} % exponentielle
\newcommand{\eps}{\varepsilon} % epsilon
\newcommand{\db}[1]{\overline{\overline{#1}}} % double barre (dyadic)
\graphicspath{{./images/},{./},{../images/}}

\begin{document}

\title{Photonic analogues of the Haldane and Kane-Mele models}
\author{Sylvain Lanneb\`{e}re\textsuperscript{1}}
\author{M\'{a}rio G. Silveirinha\textsuperscript{1,2}}
\email{To whom correspondence should be addressed:
mario.silveirinha@co.it.pt}
 \affiliation{\textsuperscript{1}
Department of Electrical Engineering, University of Coimbra and
Instituto de Telecomunica\c{c}\~{o}es, 3030-290 Coimbra, Portugal}
\affiliation{\textsuperscript{2}University of Lisbon -- Instituto
Superior T\'ecnico, Department of Electrical Engineering, 1049-001
Lisboa, Portugal}

\begin{abstract}
The condensed matter Haldane and Kane-Mele models revolutionized the
understanding of what is an ``insulator'', as they unveiled novel
classes of media that behave as metals near the surface, but are
insulating in the bulk. Here, we propose exact electromagnetic
analogues of these two influential models relying on a photonic
crystal implementation of ``artificial graphene'' subject to an
effective magnetic field. For the Haldane model, the required
effective magnetic field for photons can be emulated with a
spatially variable pseudo-Tellegen response. For the Kane-Mele
model, the spin-orbit coupling can be mimicked using matched
anisotropic dielectrics with identical permittivity and
permeability, without requiring any form of bianisotropic couplings.
Using full wave numerical simulations and duality theory we verify
that the nontrivial topology of the two proposed platforms results
in the emergence of topologically protected gapless edge states at
the interface with a trivial photonic insulator. Our theory paves
the way for the emulation of the two condensed matter models in a
photonic platform, and determines another paradigm to observe
topologically protected edges-states in a fully reciprocal
all-dielectric and non-uniform anisotropic metamaterial.
\end{abstract}

\maketitle

\section{Introduction}  
In recent years the advent of topological methods in
electromagnetism brought new perspectives for robust waveguiding
largely immune to fabrication imperfections
\cite{lu_topological_2014,lu_topological_2016,Haldane_Nobel_2017,khanikaev_two-dimensional_2017,sun_two-dimensional_2017,ozawa_topological_2018,ma_guiding_2015,ma_scattering-free_2017}.
Nontrivial photonic topological materials with a broken (non-broken)
time-reversal symmetry were shown to enable unidirectional
(bidirectional) and reflectionless edge state propagation at the
interface with a trivial electromagnetic insulator. In addition to
offering inherent optical isolation, these new paradigms to guide
light without any backscattering may also have far-reaching
repercussions in quantum optics \cite{barik_topological_2018}, in
the realization of high-efficiency lasers
\cite{harari_topological_2018,bandres_topological_2018}, or in light
harvesting \cite{topological_sink_David}.

Historically, the theory of topological photonics was largely
inspired by its electronic counterpart
\cite{raghu_analogs_2008,haldane_possible_2008}, i.e., by the
properties of electronic phases of matter whose study started in the
1980s shortly after the discovery of the integer quantum Hall effect
\cite{hasan_textitcolloquium_2010,shen_topological_2012}. In
particular, two condensed-matter models played a major role in the
development of this field: the Haldane model demonstrating that a
broken time-reversal symmetry is the key ingredient to obtain a
quantized Hall conductivity \cite{haldane_model_1988}, and the
Kane-Mele model that showed that the spin-orbit coupling may imitate
the effect of a magnetic field in time-reversal invariant electronic
systems \cite{kane_quantum_2005,kane_$z_2$_2005}. Surprisingly,
despite the many analogies drawn in the past years between
electronic and photonic systems, there are no strict electromagnetic
equivalents of the two models, and only a few first order
approximations of the Kane-Mele model were identified in
\cite{khanikaev_photonic_2013,slobozhanyuk_three-dimensional_2017}
relying on bianisotropic materials with $\Omega$-coupling.

Here, building on a recent proposal for an electronic implementation
of the Haldane model in ``artificial graphene'' made of a patterned
2D Electron Gas (2DEG)  \cite{lannebere_link_2018}, we  aim to fill
this gap and propose exact analogues of the Haldane and Kane-Mele
models in electromagnetics. To this end, in the first part of the
article, we use an analogy between the Schr\"odinger and Maxwell
equations to introduce a novel implementation of ``photonic
graphene" based on a honeycomb lattice of dielectric cylinders
embedded in a metallic background. It is shown that the magnetic
field of the Haldane model can be effectively implemented with a
spatially varying pseudo-Tellegen coupling. In this manner, we
construct an exact photonic analogue of the Haldane model. In the
second part of the article, we demonstrate that by matching the
electric and magnetic responses one obtains a precise analogue of
the Kane-Mele model in the same nonreciprocal photonic platform.
Next, it is shown that this nonreciprocal system can be linked
through a duality transformation with another all-dielectric
platform. Thereby, it follows that the Kane-Mele model can be
implemented with a fully reciprocal metamaterial made of spatially
dependent anisotropic dielectrics with no magnetoelectric coupling.
Notably, the proposed system turns out to be parity-time-duality
($\mathcal{PTD}$) symmetric \cite{silveirinha_PTD_2017}, a symmetry
that guarantees bi-directional scattering-immune propagation under
some conditions. Furthermore, our photonic implementation of the
Kane-Mele model is also found to be related to the pseudo-magnetic
field introduced by Liu and Li in \cite{liu_gauge_2015}. Thus, our
analysis unveils a rather fundamental link between time-reversal
invariant topological matter \cite{khanikaev_photonic_2013},
pseudo-magnetic fields \cite{liu_gauge_2015} and $\mathcal{PTD}$
symmetric systems \cite{silveirinha_PTD_2017,
chen_symmetry-protected_2015}.

\section{Electromagnetic Haldane model} \label{sec:Haldane_electromagn}

The Haldane model is a condensed-matter model of a spinless electron
that consists in the generalization of the tight-binding Hamiltonian
of graphene to systems with a broken inversion symmetry (IS) and/or
a broken time-reversal symmetry (TRS) \cite{haldane_model_1988,
kim_realizing_2017}. In this model the topology of
the bands --characterized by the electronic Chern number $\nu$-- is
determined by the dominant broken symmetry: a dominant broken IS has
a trivial topology with $\nu=0$ whereas a dominant broken TRS leads
to a non-trivial topology with $\nu=\pm 1$, and is characterized by
the presence of unidirectional edge states protected against
backscattering (a quantized Hall phase) at the interface with a
trivial insulator. For more details about this model the reader is
referred to \cite{haldane_model_1988,lannebere_link_2018}.

Here, we propose an electromagnetic equivalent of this
condensed-matter system relying on an analogy between the 2D
Schr\"odinger and Maxwell equations. The starting point of this
analogy is the microscopic Schr\"odinger equation that emulates the
Haldane model in artificial graphene \cite{lannebere_link_2018}, and
whose main features are summarized in  the next subsection.

\subsection{Electronic Haldane model in a 2D Electron Gas} \label{sec:Haldane_electronic}
\begin{figure}[!ht]
\centering
\includegraphics[width=.95\linewidth]{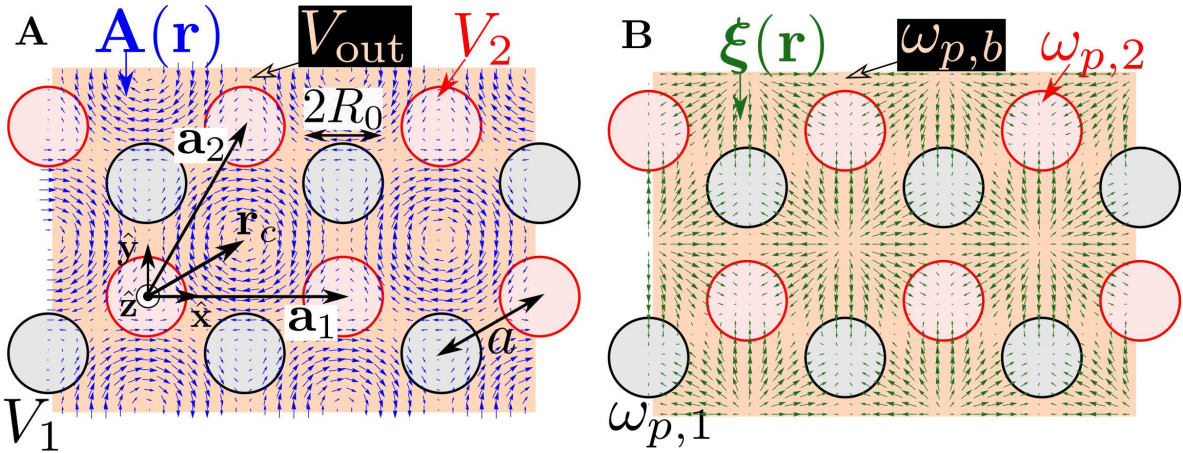}
         \caption{Schematic of (A) the electronic and (B) the equivalent photonic structure used to implement the Haldane model.
         The $V_i$'s and $\omega_{p,i}$'s with $i=1,2$ are respectively the electrostatic potentials and plasma frequencies of the materials associated with the two different sublattices,
         $V_\text{out}$ and $\omega_{p,b}$ are the electrostatic potential and the plasma frequency of the background material,
         and $\vec{A}$ and $\boldsymbol{\xi}$ are the spatially dependent magnetic vector potential and pseudo-Tellegen
         vector given by Eqs. \eqref{E:vecA} and \eqref{E:vec_xi_Haldane_photonic}, respectively.}
         \label{fig:Haldane_model}
\end{figure}

It was recently shown in \cite{lannebere_link_2018} that the Haldane
model can be implemented in artificial graphene, i.e., an electronic
platform that mimics the properties of graphene
\cite{polini_artificial_2013}, for example a  2DEG under the
influence of a periodic electrostatic potential $V(\vec{r})$ with
the honeycomb symmetry
\cite{gibertini_engineering_2009,lannebere_effective_2015,wang_observation_2018}.
As depicted in Figure \ref{fig:Haldane_model}A, a broken IS in
artificial graphene can be realized with different potentials $V_1$
and $V_2$ in each sublattice, while a broken TRS can be obtained by
applying a space-varying static magnetic field
$\vec{B}(\vec{r})=\nabla \times \vec{A}$ with zero spatial average.
As proven in \cite{lannebere_link_2018}, a magnetic vector potential
$\vec{A}$ of the form
\begin{equation} \label{E:vecA}
\vec{A}(\vec{r}) = \frac{3 B_0 a^2}{ 16 \pi^2 } \left[ \vec{b}_1
\sin\left( \vec{b}_1 \cdot \vec{R} \right) + \vec{b}_2 \sin\left(
\vec{b}_2 \cdot \vec{R} \right) + \left( \vec{b}_1 + \vec{b}_2
\right) \sin\left( \left[ \vec{b}_1 + \vec{b}_2 \right] \cdot
\vec{R} \right) \right]   \times \hat{\vec{z}}
\end{equation}
has the required symmetries to reproduce the Haldane model such that
only the phase with dominant broken TRS leads to a non-trivial
topology with $\nu\neq 0$. In the above formula $a$ is the distance
between nearest sites (scattering centers) in the honeycomb lattice,
$B_0$ is the peak magnetic field in Tesla,
$\vec{R}=\vec{r}-\vec{r}_c$ where $\vec{r}_c$ determines the
coordinates of the honeycomb cell's center and the $\vec{b}_i$'s
with $i=1,2$ are the reciprocal lattice primitive vectors.
The electronic system of Figure \ref{fig:Haldane_model}A is
characterized by a microscopic Hamiltonian
\begin{equation}
\Ham_\text{mic} = \frac{1}{2m_b}\left(\hat{\vec{p}} + e
\vec{A}(\vec{r}) \right)^2+ V(\vec{r}),
\end{equation}
where $m_b$ is the electron effective mass,
$\hat{\vec{p}}=-i\hbar\nabla$ the momentum operator and $e>0$ is the
elementary charge. It follows that the stationary states $\psi$ with
energy $E$ i.e., the solutions of the time-independent
Schr\"odinger's equation $\Ham_\text{mic}\psi=E \psi$, satisfy
\begin{align}\label{E:Schrodinger_Haldane_graphene}
 \left[ \frac{-\hbar^2}{2m_b}\left( \nabla + i \frac{e  }{\hbar} \vec{A}(\vec{r})\right)^2 + V(\vec{r}) - E \right] \psi =0.
 \end{align}
Crucially, this microscopic equation with a magnetic vector
potential $\vec{A}$ given by Eq. \eqref{E:vecA} reproduces the
symmetries of the Haldane model \cite{lannebere_link_2018}, and as
shown next, can also be used to develop an analogy with the Maxwell
equations.

\subsection{Photonic analogue} \label{sec:Haldane_electromagn_Tellegen}
The strategy adopted by us to obtain a photonic equivalent of the
Haldane model \eqref{E:Schrodinger_Haldane_graphene} is to exploit a
formal analogy between the 2D Schr\"odinger and Maxwell equations.
For the sake of clarity, the analysis is divided into two steps:
first it is shown how to create an electromagnetic equivalent of
artificial graphene in a 2D photonic crystal. Then it is explained
how to break the fundamental symmetries of the system in order to
implement the electromagnetic Haldane model.

Using a six-vector formalism for the representation of the
electromagnetic fields \cite{serdyukov_electromagnetics_2001}, the
Maxwell equations with current-sources can be written in a compact
manner as
\begin{align} \label{E:Maxwell_eqs}
 \begin{pmatrix} 0 & i
\nabla \times \vec{1}_{3\times3} \\ -i\nabla \times
\vec{1}_{3\times3} & 0 \end{pmatrix} \cdot \vec{F} = \omega \vec{M}
\cdot \vec{F} + i\vec{J},
\end{align}
where $\vec{F}=\left(\vec{E} \quad \vec{H} \right)^T$ and
$\vec{J}=\left(\vec{j}_e \quad \vec{j}_m \right)^T$ stand for
six-vectors whose components are the electric and magnetic fields
and current densities respectively, and $\vec{M}$ is the material
matrix given by
\begin{align}\label{E:matrixM}
\vec{M}(\vec{r})=\begin{pmatrix} \eps_0\db{\eps}(\vec{r}) &
\dfrac{1}{c} \db{\xi}(\vec{r})\\ \dfrac{1}{c} \db{\zeta}(\vec{r}) &
\mu_0\db{\mu}(\vec{r})
\end{pmatrix},
\end{align}
where $\db{\eps}$, $\db{\mu}$, $\db{\xi}$ and $\db{\zeta}$ are the
relative permittivity, permeability and magnetoelectric tensors
respectively. In addition, it is also assumed that the system is
invariant to translations along the $z$ direction
$\left(\frac{\partial}{\partial z}=0\right)$.

\subsubsection{Photonic artificial graphene}
During the last decade many photonic equivalents of graphene were
put forward, notably relying on photonic crystals with either
honeycomb or triangular symmetry with dielectric
\cite{raghu_analogs_2008,sepkhanov_extremal_2007,peleg_conical_2007,ochiai_photonic_2009,zandbergen_experimental_2010,bravo-abad_enabling_2012,rechtsman_photonic_2013,plotnik_observation_2014}
or metallic scatterers
\cite{han_dirac_2009,bittner_observation_2010,bittner_extremal_2012},
or alternatively with semiconductor cavities
\cite{jacqmin_direct_2014}. In the following, we suggest another
approach to engineer ``photonic graphene'' based on a direct analogy
between the 2D Schr\"odinger and Maxwell equations. To begin with,
we consider a 2D non-magnetic photonic crystal described by the
relative permittivity and permeability tensors
\begin{align}
\db{\eps}&=\eps_\parallel \left( \hat{\vec{x}}\hat{\vec{x}} + \hat{\vec{y}}\hat{\vec{y}} \right)+ \eps_{zz} \hat{\vec{z}}\hat{\vec{z}} \label{E:permit_artificial_graphene}\\
\db{\mu}&=\mathbf{1}_{3\times3}.
\label{E:permeab_artificial_graphene}
\end{align}
In the absence of current-sources ($\vec{J}=0$) the wave equation
for transverse electric (TE) polarized waves ($\vec{E}=E_z
\hat{\vec{z}}$) is given by Eq. \eqref{E:wave_equation_TE} of the
Appendix with $\boldsymbol{\xi}=0$. Supposing that the
$zz$-component of the permittivity tensor is described by a Drude
dispersion model, $\eps_{zz}=1-\frac{\omega_p^2 }{\omega^2}$, the
wave equation \eqref{E:wave_equation_TE} reduces to
\begin{align}  \label{E:Maxwell_artificial_graphene_TE}
\left[ \nabla^2 + \frac{\omega^2}{c^2}    -
\frac{\omega_p^2}{c^2}\right]  E_z    &=     0.
\end{align}
By comparing this formula with the Schr\"odinger equation of
electronic artificial graphene, i.e., Eq. \eqref{E:Schrodinger_Haldane_graphene} in the absence of magnetic
field ($\vec{A}=0$), it is seen that the solutions of both equations
can be matched by taking
\begin{align}
\frac{2 m_b E }{\hbar^2} &= \frac{\omega^2}{c^2},  \label{E:equival_E_omega}\\
\frac{2m_b V(\vec{r})}{\hbar^2}   &= \frac{\omega_p^2(\vec{r})}{c^2}
\label{E:equival_V_omegaP}.
\end{align}
In order that the analogy is perfect the plasma frequency $\omega_p$
of the material is required to be spatially dependent (a photonic
crystal) with the same periodicity as the electronic potential.
Incidentally it can be noted that because the right-hand side of
\eqref{E:equival_V_omegaP} is always positive, this equivalence is
only possible for a positive electric potential $V$. However because
only the difference $V-E$ is relevant in the Schr\"odinger equation
\eqref{E:Schrodinger_Haldane_graphene}, it is always feasible to
transform a negative electric potential into a positive one by
adding an overall constant potential $U$ to both $V$ and $E$, such
that the potential is transformed as $V \to V+ U >0$ and the origin
of energy is shifted as $E \to E +U$. Thus, the relations
\eqref{E:equival_E_omega} and  \eqref{E:equival_V_omegaP} can always
be satisfied when the electric potential has a lower bound. Then,
following
\cite{gibertini_engineering_2009,lannebere_effective_2015,wang_observation_2018},
we conclude that a strict photonic equivalent of artificial graphene
can be implemented in a photonic crystal made of dielectric
cylinders (``potential wells'' with $\omega_{p,i} = 0, i=1,2$)
arranged in a honeycomb lattice in a metallic background (with
$\omega_{p,b}
> 0$). To our best knowledge, this is the first proposal of a
photonic equivalent of graphene based on a photonic crystal with a
metallic background.

\subsubsection{Magnetic field for photons with a spatially variable pseudo-Tellegen response}
Now that we identified a photonic system with the same properties as
graphene, the second step to emulate the Haldane model is to break
the fundamental IS and TRS symmetries in this platform. In the
electronic model \eqref{E:Schrodinger_Haldane_graphene} a broken IS
is obtained by using different values of $V_1$ and $V_2$ in each
sublattice of the honeycomb structure (Figure
\ref{fig:Haldane_model}A). From the equivalence relation
\eqref{E:equival_V_omegaP}, the same effect may be attained in
photonics by using scatterers with different plasma frequencies
$\omega_{p,1}$, $\omega_{p,2}$ in each sublattice, as depicted in
Figure \ref{fig:Haldane_model}B.

On the other hand, a broken TRS is generally trickier to implement
for photons. Here we create an effective magnetic field for photons
by taking advantage of a bisanisotropic response of the medium. In
particular, in addition to the effective permittivity
\eqref{E:permit_artificial_graphene} and permeability
\eqref{E:permeab_artificial_graphene}, it is assumed that the
material response is nonreciprocal with a symmetric pseudo-Tellegen
response (following the classification of
\cite{serdyukov_electromagnetics_2001}) with the magnetoelectric
coupling tensors given by:
\begin{align}\label{E:magnetoelectric_coupling_Haldane}
\db{\xi}&=\db{\zeta}  =   \boldsymbol{\xi} \otimes \hat{\vec{z}}
+\hat{\vec{z}} \otimes \boldsymbol{\xi},
\end{align}
where $\boldsymbol{\xi}=\xi_x  \hat{\vec{x}}+ \xi_y  \hat{\vec{y}}$
is a generic vector lying in the $xoy$ plane. Notably as
demonstrated in appendix \ref{sec:general_wave_eq_pseudo-Tellegen},
such a magnetoelectric coupling does not mix the TE and TM
polarizations. In particular, from \eqref{E:wave_equation_TE} it
follows that the wave equation for TE-polarized waves is:
\begin{align}  \label{E:Maxwell_Haldane_graphene_TE}
\left[\left( \nabla   -  i  \frac{\omega}{c}   \hat{\vec{z}} \times
\boldsymbol{\xi}  \right)^2 + \frac{\omega^2}{c^2}    -
\frac{\omega_p^2}{c^2}\right]  E_z    &=     0.
\end{align}
By comparing this equation with the microscopic electronic Haldane
model \eqref{E:Schrodinger_Haldane_graphene} it is seen that, in
addition to the equivalence relations \eqref{E:equival_E_omega} and
\eqref{E:equival_V_omegaP}, the solutions of both equations can be
matched by considering a spatially dependent pseudo-Tellegen
response such that
\begin{align}
\frac{e  }{\hbar} \vec{A}(\vec{r}) &=  \frac{\omega}{c}
\boldsymbol{\xi}(\vec{r})\times \hat{\vec{z}}
\label{E:equival_A_xi}.
\end{align}
Strikingly this relation proves that in two-dimensional scenarios a
spatially varying pseudo-Tellegen response
$\boldsymbol{\xi}(\vec{r})$ is the equivalent for photons of a
magnetic field acting on electrons. Thus, it follows that Eq. \eqref{E:Maxwell_Haldane_graphene_TE} with a Tellegen-type coupling
determined by \eqref{E:equival_A_xi} is the exact photonic
counterpart of Eq. \eqref{E:Schrodinger_Haldane_graphene} and
thereby yields a photonic Haldane model. Here, we note that Ref.
\cite{jacobs_photonic_2015} studied a topological system with a
similar Tellegen-coupling, but which is not an analogue of the
Haldane model. Furthermore, Ref. \cite{he_photonic_2016}
investigated a topological photonic crystal with an anti-symmetric
moving-type nonreciprocal coupling
\cite{serdyukov_electromagnetics_2001,kong_electromagnetic_2000},
which is different from the symmetric-Tellegen response considered
by us.

By substituting the magnetic vector potential
\eqref{E:vecA} into \eqref{E:equival_A_xi}, the spatially dependent
pseudo-Tellegen response is found to be given by
\begin{align} \label{E:vec_xi_Haldane_photonic}
\boldsymbol{\xi}(\vec{r})    &= \xi_0 \frac{ \sqrt{3}a}{ 4 \pi
}\left[ \vec{b}_1  \sin\left( \vec{b}_1 \cdot \vec{R} \right) +
\vec{b}_2 \sin\left( \vec{b}_2 \cdot \vec{R} \right) + \left(
\vec{b}_1 + \vec{b}_2 \right) \sin\left( \left[ \vec{b}_1 +
\vec{b}_2 \right] \cdot \vec{R} \right) \right],
\end{align}
where $\xi_0$ is the dimensionless peak amplitude of the
pseudo-Tellegen vector, which is linked to the parameters of the
original electronic system as
\begin{align} \label{E:xi_0}
\xi_0   &= \frac{\sqrt{3} }{ 4 \pi } \frac{e c B_0 a}{
\hbar\omega}.
\end{align}
For given values of $B_0$ and $a$ (corresponding to a specific
implementation of the electronic Haldane model) the peak amplitude
$\xi_0$ is $\omega$-dependent. Importantly, the reality condition
for the material matrix $\vec{M}(\omega)$ imposes that
$\boldsymbol{\xi}(\omega)=\boldsymbol{\xi}^\ast(-\omega)$
\cite{agranovich_crystal_1984}, a condition  that is obviously
violated by \eqref{E:xi_0}. For simplicity, we shall assume in this
study that $\xi_0$ is frequency independent. A constant $\xi_0$ is
equivalent to a nontrivial magnetic field $B_0$ whose strength is
frequency dependent. Evidently, for positive frequencies the sign of
$\xi_0$ is the same as the sign of $B_0$, and hence the topological
phases of the relevant photonic and electronic systems are strictly
linked. Furthermore, provided  Eq. \eqref{E:xi_0} is satisfied for
some frequency in the band gap, the physics of the two systems is
essentially the same in the spectral region of the gap. Finally, we
note that the proposed photonic platform emulates precisely the
Haldane model even if the frequency dependence of $\xi_0$ does not
follow \eqref{E:xi_0}. Indeed, independent of the dispersion of
$\xi_0$, in a tight-binding binding approximation the system is
rigorously described by Haldane's theory.

In summary, it was demonstrated that a bianisotropic metamaterial
characterized by effective parameters of the form
\eqref{E:permit_artificial_graphene},
\eqref{E:permeab_artificial_graphene},
\eqref{E:magnetoelectric_coupling_Haldane} and such that $\eps_{zz}$
is given by a Drude model and $\boldsymbol{\xi}$ by equation
\eqref{E:vec_xi_Haldane_photonic} yields an analogue of the Haldane
model for photons. The phase diagram relating the
photonic gap Chern number $\mathcal{C}_\text{gap}$ to the strength of the
pseudo-Tellegen response and to the asymmetry of the sub-lattices
(measured by $\omega_{p,2}-\omega_{p,1}$) can be found from the
corresponding electronic phase diagram (Figure 1(d) of
\cite{lannebere_link_2018}) using the relations
\eqref{E:equival_E_omega}, \eqref{E:equival_V_omegaP},
\eqref{E:equival_A_xi} and \eqref{E:xi_0}.

\subsection{Numerical examples} \label{sec:num_ex_Haldane}
To illustrate the ideas developed so far, next we present the band
diagrams and the typical field profiles associated with each
topological phase of the photonic Haldane model. All the simulations
presented here were numerically  obtained with a dedicated
finite-differences frequency-domain method (FDFD) whose details can
be found in \cite{lannebere_effective_2015}.

The electromagnetic system is constructed from a dual electronic
system through the equivalence relations \eqref{E:equival_E_omega},
\eqref{E:equival_V_omegaP} and \eqref{E:equival_A_xi}. We choose the
original electronic platform as the patterned 2DEG of Figure
\ref{fig:Haldane_model}A with parameters $V_1=V_2=0$,
$V_\text{out}\frac{m_b a^2}{\hbar^2} \approx 15.83$, $\vec{A}=0$,
$R_0/a=0.35$ and  $m_b=0.067m$, with $m$ the electron rest mass.
With such parameters the electronic system behaves as graphene near
the normalized energy $E\frac{m_b a^2}{\hbar^2} \approx 9.4 $ where
its band diagram consists of two Dirac cones centered at the
high-symmetry $K$ and $K'$ points. For $a=150 ~\nano\meter$ the
system reduces to the artificial graphene studied in
\cite{lannebere_link_2018,lannebere_effective_2015} after adding a
constant potential $U=0.8~\milli\electronvolt$ to the whole
structure to guarantee that $V\geq0$.

\begin{figure}[!ht]
\centering
\includegraphics[width=.95\linewidth]{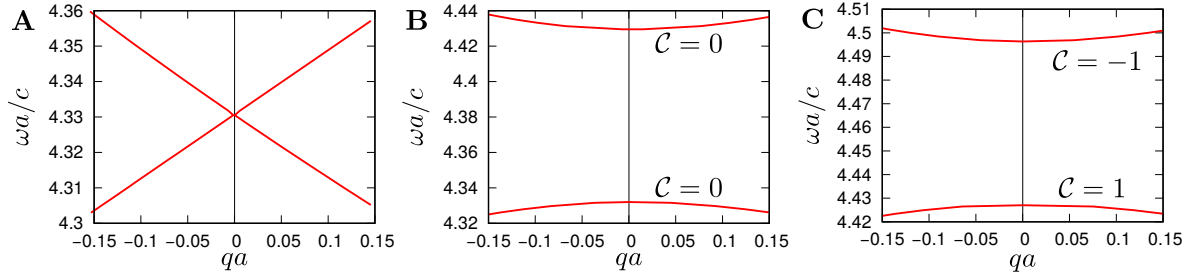}
         \caption{Normalized dispersion diagrams for the photonic crystal of Figure \ref{fig:Haldane_model}B with
          $\omega_{p,b} \frac{a}{c}\approx 5.63 $. (A) ``photonic graphene'' with $\omega_{p,1}=\omega_{p,2}= 0$  and $\boldsymbol{\xi}=0$.
          (B) ``photonic graphene'' with a broken IS such that $\boldsymbol{\xi}=0$, $\omega_{p,1}=0$ and $\omega_{p,2}\frac{a}{c}\approx 1.09 $.
          (C) ``photonic graphene'' with a broken TRS such that $\xi_0\approx 0.677$  and $\omega_{p,1} =\omega_{p,2}=0$.
          The photonic Chern numbers $\mathcal{C}$ associated with the two bands are given in insets.}
         \label{fig:BD_art_graph}
\end{figure}

The direct application of the equivalence relations
\eqref{E:equival_E_omega}, \eqref{E:equival_V_omegaP} to the
artificial graphene leads to a simple photonic crystal made of air
cylinders ($\omega_{p,1}=\omega_{p,2}=0$) embedded in a metal with a
Drude dispersion. The band diagram of the photonic crystal close to
the Dirac point $K$ is depicted in Figure \ref{fig:BD_art_graph}A as
a function of the wavevector $q=|\vec{k}-\vec{K}|$ taken relatively
to the $K$ point.
As expected,  the frequency dispersion is approximately a linear
function of $q$ forming a Dirac cone at $K$ and $K'$ (not shown),
thus validating that the proposed structure is a photonic equivalent
of graphene. As seen in Figures \ref{fig:BD_art_graph}B and C, by
breaking either the IS ($\omega_{p,1}\neq\omega_{p,2}$) or the TRS
($\boldsymbol{\xi}\neq 0$) it is possible to open a band-gap in the
band diagram. Moreover, similar to the electronic Haldane model, the
phases induced by each of the broken symmetries are topologically
distinct, as expressed by the different values of the photonic Chern
number $\mathcal{C}$. Specifically, $\mathcal{C}=0$ for the phase
with broken IS whereas $\mathcal{C}=\pm1$ for the phase with broken
TRS. It should be noted that with a frequency independent $\xi_0$
and with $\omega_p a/c = \mathit{const.}$ the properties of the photonic
system are fully scalable with frequency, and the exact gap spectral
range is determined only by $a$.

\begin{figure}[!ht]
\centering
\includegraphics[width=.95\linewidth]{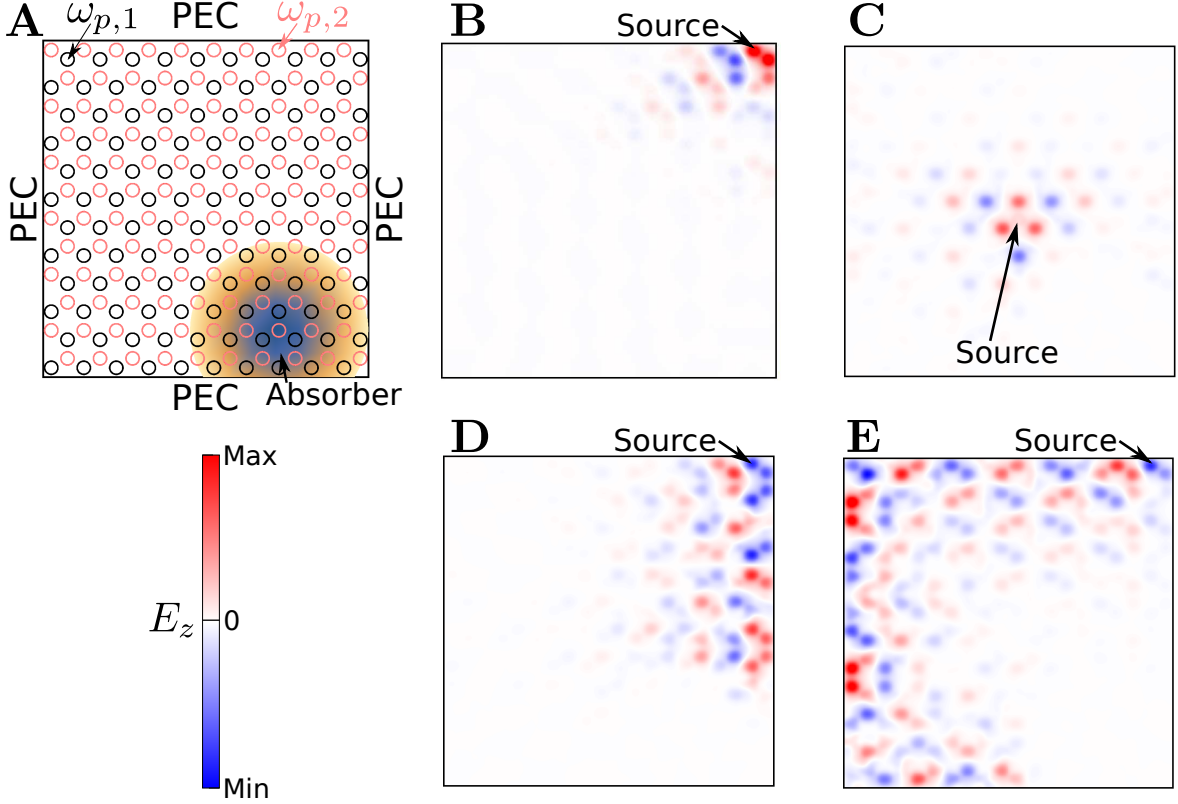}
         \caption{(A) Schematic of the simulated cavity: the photonic crystal of Figure \ref{fig:Haldane_model}B with an absorber located at the right-bottom region
          and surrounded by PEC walls. The pseudo-Tellegen response $\boldsymbol{\xi}$ is not represented in the figure.
          (B) Time snapshot ($t=0$) of $E_z$ for the material with a broken IS of Figure \ref{fig:BD_art_graph}B.
          (C)--(E) Time snapshot ($t=0$) of $E_z$ for the material with a broken
          TRS of Figure \ref{fig:BD_art_graph}C. The pseudo-Tellegen parameter is $\xi_0\approx 0.677$ in (C)--(D) and $\xi_0\approx -0.677$ in (E).}
         \label{fig:edge_states}
\end{figure}

According to the bulk-edge correspondence
\cite{lu_topological_2014,raghu_analogs_2008,silveirinha_proof_2018}
the difference in the Chern numbers should manifest itself directly
on the propagation of the edge states at the interface with a
trivial photonic insulator. In particular, it is expected that the
phase with broken TRS symmetry supports unidirectional edge states
protected against backscattering with a dispersion that spans the
entire band-gap. To confirm this property the solutions $E_z$ of the
wave equation in the closed cavity of Figure \ref{fig:edge_states}A,
\begin{align}
\left[\left( \nabla   -  i  \frac{\omega}{c}   \hat{\vec{z}} \times
\boldsymbol{\xi}  \right)^2 + \frac{\omega^2}{c^2}    -
\frac{\omega_p^2}{c^2}\right]  E_z     &=   i \omega \mu_0 j_e,
\end{align}
were computed with the FDFD method \cite{lannebere_effective_2015}
with the wave excited by a point-like electric current distribution
$j_e$. The cavity walls were chosen to be perfect electric
conductors (PEC), i.e., the Haldane photonic crystal is surrounded
by a trivial insulator. The oscillation frequency of the current
source is centered in the band-gap.
To ease the field visualization in the closed cavity, an absorber
was placed at the right-bottom part of the structure. The absorption
is stronger near the center (darker colors in Figure
\ref{fig:edge_states}A imply a stronger absorption).

The excited field profiles for the distinct topological phases of
photonic Haldane graphene are represented in Figure
\ref{fig:edge_states}B--E.
As seen in Figure \ref{fig:edge_states}B, the phase characterized by
a broken IS does not support any bulk or unidirectional edge mode at
this frequency, confirming that this phase is topologically trivial.
The absence of bulk modes for the phase with a broken TRS is also
verified in Figure \ref{fig:edge_states}C, which reveals that for an
excitation far from the edges, the fields decay rapidly with the
distance to the source.

The situation is dramatically different for sources positioned near
the boundaries. Figures \ref{fig:edge_states}D and E show that in
such a case an unidirectional edge state with propagation direction
locked to the sign of $\xi_0$ is excited near the PEC walls.
Crucially, the topologically protected modes do not experience any
backscattering at the corners of the cavity. 
\begin{figure}[!ht]
\centering
\includegraphics[width=.45\linewidth]{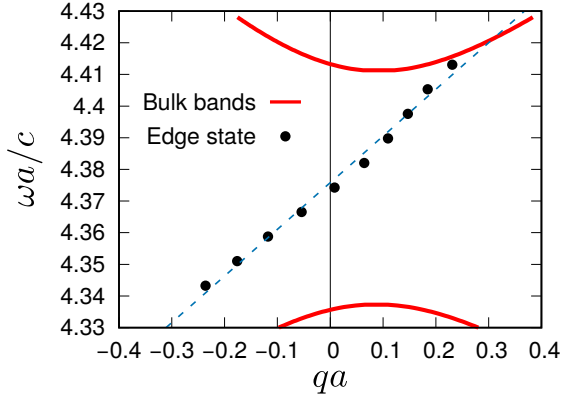}
         \caption{Dispersion diagram of the bulk and edge modes (top interface) as a function of the normalized wavevector.
         The structure parameters are as in Figure \ref{fig:BD_art_graph}C.
         The slight asymmetry of the bulk bands with respect to $q=0$ and the frequency shift with respect to Figure \ref{fig:BD_art_graph}C are numerical artifacts
         caused by a relatively coarse mesh in the simulations of the electrically large topological cavity of Figure \ref{fig:edge_states}.}
         \label{fig:fit_edge_states}
\end{figure}
The edge mode dispersion in the band-gap region was found by
numerically fitting the spatial variation of the edge waves to that
of a Bloch wave. As depicted in Figure \ref{fig:fit_edge_states},
the edge mode dispersion is approximately linear and for the edge
wave propagating attached to the top interface it is centered about
the $K$ point and spans the entire band gap. The gap Chern number
--given by the sum of the Chern numbers of the bands below the gap,
including the negative frequency bands-- is
$\mathcal{C}_\text{gap}=\mathrm{sgn}(\xi_0)$ for the phase with a
dominant broken TRS. In agreement with the results of
\cite{silveirinha_quantized_2019, silveirinha_proof_2018} the gap
Chern number is positive (negative) for an energy flow in the
clockwise (anticlockwise) direction. The electric field time animations of the examples of Figure \ref{fig:edge_states}D and E are available in the Supplementary  Material.

It is relevant to mention that because of computational restrictions
the maximum number of mesh cells allowed by our numerical code is
limited. For this reason the results of Figures
\ref{fig:edge_states} and \ref{fig:fit_edge_states} -- for a
computational domain formed by many cells -- were obtained with a
relatively low mesh density. For consistency, the band structure
shown in Fig. \ref{fig:fit_edge_states} was calculated using the
same mesh as the one used to obtain the edge states dispersion. The
coarser mesh leads to a slight numerical shift with respect to the
more exact band structure results of Figure \ref{fig:BD_art_graph}.

Also because of computational limitations, the calculation of the
edge modes dispersion is challenging when they are weakly confined
to the cavity walls. To obtain the most accurate results, we focused
on a phase with a broken TRS, but with a preserved inversion
symmetry ($\omega_{p,1}=\omega_{p,2}=0$) to maximize the bandgap
width. Indeed, the edge state confinement is stronger for a larger gap
energy. We numerically verified that as long as TRS remains the
dominant broken symmetry the introduction of a broken IS
($\omega_{p,1}\neq\omega_{p,2}$) does not affect the topological
properties of the system, and the presence of gapless edge states.

In conclusion, we numerically demonstrated that the structure of
Figure \ref{fig:Haldane_model}B described by Eq. \eqref{E:Maxwell_Haldane_graphene_TE} is a photonic equivalent of
the Haldane model, with topological properties determined by the
dominant broken symmetry exactly as its electronic counterpart.

%%%%%%%%%%%%%%%%%%%%%%%%%%%%%%%%%%%%%%%%%%%%%%%%%%%%%%%%%%%%%%%%%%%%%%%%%%%%%%%%%%%%%%%%%%%%%%%%%%%%%%%%%
%%%%%%%%%%%%%%%%%%%%%%%%%%%%%%%%%%%%%%%%%%%%%%%%%%%%%%%%%%%%%%%%%%%%%%%%%%%%%%%%%%%%%%%%%%%%%%%%%%%%%%%%%
\section{Photonic Kane-Mele model} \label{sec:Kane-Mele_electromag}

The electronic Kane-Mele model was introduced around 2005 in a
series of two papers \cite{kane_quantum_2005,kane_$z_2$_2005}. It
describes a time-reversal invariant mechanism that effectively
imitates the effect of a magnetic field for each electron spin. The
Kane-Mele model is also based on the tight-binding Hamiltonian of
graphene but with the two different spins coupled by the spin-orbit
interaction. The most remarkable prediction of this model is that
the spin-orbit coupling can induce topologically nontrivial band
gaps characterized by the presence of scattering-immune
spin-polarized edge currents, a phenomenon known as the quantum spin
Hall effect. This effect was first experimentally observed in HgCdTe
quantum wells \cite{konig_quantum_2007}. Recently, photonic analogues
of such non-trivial electronic edge states were studied in several
electromagnetic systems with the spin degree of freedom mimicked by
the light polarization and the spin-orbit coupling by, for example,
a bianisotropic coupling
\cite{khanikaev_photonic_2013,slobozhanyuk_three-dimensional_2017}
or by particular symmetries of the waveguide
\cite{silveirinha_PTD_2017,chen_symmetry-protected_2015, martini_exact_2019}.

Mathematically, the Kane-Mele model can be as well regarded as two
copies of the Haldane model with each electron spin experiencing an
opposite magnetic field
\cite{shen_topological_2012,kane_$z_2$_2005}. By adopting this point
of view and building on the results of the previous section, we show
in Sect. \ref{sec:Kane-Mele_nonreciprocal} how to mimic the Kane-Mele
model with a nonreciprocal photonic platform with a pseudo-Tellegen
response. We establish a link between this platform and the class of
$\mathcal{PTD}$-invariant systems, i.e., systems invariant under the
composition of the parity $\mathcal{P}$, time-reversal
$\mathcal{T}$, and duality $\mathcal{D}$ transformations
\cite{silveirinha_PTD_2017}. Finally, in Sect.
\ref{sec:Kane-Mele_reciprocal} using duality theory we propose an
alternative implementation of the Kane-Mele model in a fully
reciprocal platform made of anisotropic dielectrics.

\subsection{Kane-Mele model in a nonreciprocal system} \label{sec:Kane-Mele_nonreciprocal}
We consider here the photonic crystal of Figure
\ref{fig:Haldane_model}B, with the same relative permittivity
\eqref{E:permit_artificial_graphene} and magnetoelectric tensors
\eqref{E:magnetoelectric_coupling_Haldane} as in Section
\ref{sec:Haldane_electromagn}, but with matched permittivity and
permeability tensors:
\begin{align}
\db{\mu} = \db{\eps}&= \eps_\parallel \left( \hat{\vec{x}}\hat{\vec{x}} + \hat{\vec{y}}\hat{\vec{y}} \right)+ \eps_{zz} \hat{\vec{z}}\hat{\vec{z}} \label{E:perms_Kane-Mele}\\
\db{\xi}&=\db{\zeta} =  \boldsymbol{\xi} \otimes \hat{\vec{z}}
+\hat{\vec{z}} \otimes \boldsymbol{\xi}.
\label{E:magnetoelec_Kane-Mele}
\end{align}
As in Section \ref{sec:Haldane_electromagn}, it is assumed that
$\eps_\parallel=\mu_\parallel$ are space independent and that the
$zz$ component of the permittivity and permeability tensors are
given by a Drude model
$\eps_{zz}(\vec{r})=\mu_{zz}(\vec{r})=1-\frac{\omega_p^2(\vec{r})
}{\omega^2}$ in the frequency range of interest. From Eqs.\eqref{E:wave_equation_TE} and \eqref{E:wave_equation_TM} of the
appendix, the wave equations for transverse electric (TE)
($\vec{E}=E_z \hat{\vec{z}}$) and transverse magnetic (TM)
($\vec{H}=H_z \hat{\vec{z}}$) polarized waves are
\begin{subequations}\label{E:Maxwell_KaneMele}
\begin{align}   \label{E:Maxwell_KaneMele_TE}
\left[\left( \nabla   -  i  \frac{\omega}{c}   \hat{\vec{z}} \times
\boldsymbol{\xi}(\vec{r})  \right)^2 +   \eps_\parallel
\left(\frac{\omega^2}{c^2}- \frac{\omega_p^2(\vec{r})
}{c^2}\right)\right]  E_z     &=     0,
\\
\left[\left( \nabla  + i  \frac{\omega}{c}     \hat{\vec{z}} \times
\boldsymbol{\xi}(\vec{r})  \right)^2   +
\eps_\parallel\left(\frac{\omega^2}{c^2}-\frac{\omega_p^2(\vec{r})
}{c^2}\right)  \right] H_z  &=0. \label{E:Maxwell_KaneMele_TM}
\end{align}
\end{subequations}
Remarkably, even though the two polarizations are uncoupled, the
wave propagation of TE and TM waves is ruled essentially by the same
equation, except for the pseudo-Tellegen vector which has an
opposite orientation for TE and TM waves. Here, $\eps_\parallel$
only plays the role of a normalization factor and can without loss
of generality be taken equal to unity. In this case and for
$\boldsymbol{\xi}$ given by Eq. \eqref{E:vec_xi_Haldane_photonic}, the wave equations
\eqref{E:Maxwell_KaneMele} reduce to two copies of the photonic
Haldane model \eqref{E:Maxwell_Haldane_graphene_TE} with an opposite
orientation of the pseudo-Tellegen vector. By virtue of Eq. \eqref{E:equival_A_xi}, this situation corresponds in the electronic
case to two copies of the Haldane model with opposite magnetic
fields, which is precisely the Kane-Mele model
\cite{kane_quantum_2005,kane_$z_2$_2005}. Hence, Eqs.
\eqref{E:Maxwell_KaneMele} yield a strict photonic analogue of the
Kane-Mele model with the spin-orbit coupling mimicked by the
nonreciprocal pseudo-Tellegen response. From Sect.
\ref{sec:Haldane_electromagn}, it is clear that the pseudo-Tellegen
coupling opens a topologically non-trivial band-gap wherein TE and
TM polarized edge states can propagate in opposite directions
without back reflections at the interface with a trivial photonic
insulator.

Interestingly, it can be readily checked that a photonic platform
with effective parameters that satisfy
\eqref{E:perms_Kane-Mele}--\eqref{E:magnetoelec_Kane-Mele} is
$\mathcal{P}\mathcal{T}\mathcal{D}$-invariant
\cite{silveirinha_PTD_2017}. The scattering phenomena in
$\mathcal{P}\mathcal{T}\mathcal{D}$-invariant systems is
characterized by an anti-symmetric scattering matrix, and this
property guarantees that a generic
$\mathcal{P}\mathcal{T}\mathcal{D}$-invariant microwave network is
matched at all ports \cite{silveirinha_PTD_2017, martini_exact_2019}. In
particular, any $\mathcal{P}\mathcal{T}\mathcal{D}$-invariant
waveguide that supports a single propagating mode (for fixed
direction of propagation) is completely immune to back-reflections
\cite{silveirinha_PTD_2017}. Note that
$\mathcal{P}\mathcal{T}\mathcal{D}$-invariant systems are always
bidirectional. Thus, the absence of backscattering in the proposed
photonic Kane-Mele model can also be explained as a consequence of
$\mathcal{P}\mathcal{T}\mathcal{D}$-invariance.

\subsection{Kane-Mele model in a reciprocal system} \label{sec:Kane-Mele_reciprocal}
While the ideas developed in the last section are interesting from
a theoretical standpoint, their impact on more practical grounds
is admittedly limited, mainly because of the difficulty to obtain a
pseudo-Tellegen response \cite{astrov_magnetoelectric_1961}. Next,
we show that the nonreciprocal $\mathcal{PTD}$-invariant platform
studied in Section \ref{sec:Kane-Mele_nonreciprocal} can be
transformed into an equivalent fully reciprocal non-bianisotropic
system by means of a duality transformation
\cite{silveirinha_PTD_2017}.

The key observation is that a duality transformation acts
exclusively on the fields, leaving the space and time unaffected
\cite{silveirinha_PTD_2017,prudencio_geometrical_2014,prudencio_optical_2016}.
Thereby, the wave phenomena and topological properties of two
systems linked by a duality transformation are fundamentally the
same. In particular a duality transformation preserves the band
diagram dispersion and the immunity to back-scattering
\cite{silveirinha_PTD_2017,prudencio_asymmetric_2015,
prudencio_geometrical_2014}.

As in \cite{silveirinha_PTD_2017}, we consider a duality
transformation $\mathcal{D}$ of the form
\begin{align} \label{E:duality_transformation_specific}
\mathcal{D} &= \frac{1}{\sqrt{2}}\begin{pmatrix}
  \mathbf{1}_{3\times3}& \eta_0 \mathbf{1}_{3\times3}\\
 -\eta_0^{-1}  \mathbf{1}_{3\times3} &   \mathbf{1}_{3\times3}
\end{pmatrix}
\end{align}
where $\eta_0=\sqrt{\mu_0/\eps_0}$ is the free space impedance. The
duality transformation changes the material parameters  as
$\vec{M}(\vec{r}) \xrightarrow{\mathcal{~D~}}
\vec{M}'(\vec{r})\equiv \det(\mathcal{D})\cdot (\mathcal{D}^{-1})^T
\cdot \vec{M}(\vec{r}) \cdot \mathcal{D}^{-1}$
\cite{silveirinha_PTD_2017}. For a bianisotropic material with
matched permittivity and permeability  $\db{\mu} = \db{\eps}$ and
matched magnetoelectric couplings $\db{\xi}=\db{\zeta}$, it leads to
a new material described by
\begin{align}
 \vec{M}' = \begin{pmatrix}
  \eps_0  \left( \db{\eps} +  \db{\xi} \right) & 0 \\ 0 &  \mu_0 \left(\db{\eps}- \db{\xi} \right)
\end{pmatrix}.
\end{align}
Notably the new material has no magnetoelectric coupling, and for
symmetric $\db{\eps}$ and $\db{\xi}$ tensors the response is
\emph{reciprocal}. If the parameters of the original material are
given by \eqref{E:perms_Kane-Mele}-\eqref{E:magnetoelec_Kane-Mele},
then the permittivity $\db{\eps}\,'$ and permeability $\db{\mu}\,'$
tensors of the new material are explicitly:
\begin{subequations}\label{E:eps_mu_dual_material}
\begin{align}
 \db{\eps}\,'(\vec{r}) &=      \begin{pmatrix}
  \eps_\parallel & 0 & \xi_x(\vec{r}) \\ 0 & \eps_\parallel & \xi_y(\vec{r}) \\ \xi_x(\vec{r}) & \xi_y(\vec{r}) & \eps_{zz}(\vec{r})
\end{pmatrix}, \label{E:eps_dual_material}\\
  \db{\mu}\,'(\vec{r}) &=  \begin{pmatrix}
 \eps_\parallel & 0 & -\xi_x(\vec{r}) \\ 0 & \eps_\parallel & -\xi_y(\vec{r}) \\ -\xi_x(\vec{r}) & -\xi_y(\vec{r}) & \eps_{zz}(\vec{r})
\end{pmatrix}. \label{E:mu_dual_material}
\end{align}
\end{subequations}
Interestingly, it can be readily verified that a system described by
the tensors $\db{\eps}\,'$ and $\db{\mu}\,'$ satisfies Eq. (12)
of \cite{silveirinha_PTD_2017} implying that it is
$\mathcal{P}\mathcal{T}\mathcal{D}$-invariant. This further confirms
that the proposed platform can support scattering-immune edge modes. Furthermore,
the tensors $\db{\eps}\,'$ and $\db{\mu}\,'$ are symmetric
indicating that the transformed system is as well reciprocal.

The tensors $\db{\eps}\,'$ and $\db{\mu}\,'$ have the same diagonal
part whereas their out-of-diagonal elements have opposite signs.
Clearly, the duality transformation preserves the diagonal elements
of the permittivity and permeability whereas the pseudo-Tellegen
response of the original material becomes the out-of-diagonal part
of $\db{\eps}\,'$ and $\db{\mu}\,'$. Evidently, the out-of-diagonal
parameters play the role of an effective magnetic field for photons
and are responsible for the bidirectional and reflectionless edge
wave propagation in this structure. Interestingly, Liu and Li have
previously shown using totally different physical arguments that
spatially dependent anisotropic media with a structure analogous to
\eqref{E:eps_mu_dual_material} can be used to create a pseudomagnetic
field for photons \cite{liu_gauge_2015,liu_polarization-dependent_2015,liu_invisibility_2017}. Thus, our theory merges the
concepts of pseudomagnetic fields and time-reversal invariant
topological matter as particular cases of the broader class of
$\mathcal{P}\mathcal{T}\mathcal{D}$-invariant systems
\cite{silveirinha_PTD_2017}.

It is highlighted that the wave propagation in a structure described
by the parameters \eqref{E:eps_mu_dual_material} is fully determined
by the wave propagation in the two associated copies of the photonic
Haldane model. 
Thereby, similar to Sec. \ref{sec:num_ex_Haldane}, the material characterized by \eqref{E:eps_mu_dual_material} supports
topologically protected (but bidirectional) edge states. The cavity walls should enforce the
($\mathcal{PTD}$-invariant) mixed boundary conditions $E_z=0$ and $H_z=0$. These ``soft'' boundary conditions 
were originally studied by Kildal and can be implemented with corrugated surfaces \cite{kildal_artificially_1990}.
The solution of the outlined cavity
problem may be strictly related through the duality transformation
\eqref{E:duality_transformation_specific} with the numerical
simulations of Sec. \ref{sec:num_ex_Haldane}.

In conclusion, we proposed a fully reciprocal
$\mathcal{PTD}$-invariant photonic platform based on non-uniform
anisotropic dielectrics that supports bidirectional edge mode
propagation protected against back-reflections. Importantly, the
concept of $\mathcal{PTD}$-invariance is a single-frequency
condition, and thereby it is sufficient that the material parameters
satisfy Eqs.\eqref{E:eps_mu_dual_material} in some
frequency in the band-gap to observe the scattering-immune
bidirectional edge state propagation. This property let us hope that
even though challenging a practical implementation of the proposed
metamaterials with spatially dependent parameters may be feasible in
the future.

\section{Conclusions}
We used an analogy between the 2D Schr\"odinger and Maxwell
equations to obtain an electromagnetic equivalent of the electronic
Haldane and Kane-Mele models. First, we introduced a novel realization of photonic graphene based on a photonic crystal
formed by dielectric cylinders arranged in a honeycomb lattice and
embedded in a metallic host with a Drude-type dispersion. Then, it
was shown that a spatially varying pseudo-Tellegen coupling is the
photonic equivalent of a magnetic field acting on electrons. Using
this result, we proposed an exact electromagnetic analogue of the
Haldane model.

Furthermore, by imposing that the permittivity and the permeability
are matched, it is possible to create two copies of the photonic
Haldane model in the same physical platform, and in this manner
implement the Kane-Mele model. Interestingly, this nonreciprocal
platform is related through a duality transformation with a much
simpler reciprocal system with the same edge states. Thereby, our
analysis proves that the Kane-Mele model can be rigorously
implemented using matched non-bianisotropic dielectrics and that
such structures can support bi-directional edge states immune to
back-scattering. The link between this system and
$\mathcal{P}\mathcal{T}\mathcal{D}$-invariant materials was
established. Furthermore, it was highlighted that all the known
mechanisms that enable the propagation of light in reciprocal
structures with no back-scattering, e.g., relying on pseudomagnetic
fields or time-reversal invariant topological insulators, fall under
the umbrella of $\mathcal{P}\mathcal{T}\mathcal{D}$-invariant
systems.

% EDITED
\section*{Acknowledgement}
This work was partially funded by the IET under the A F Harvey
Engineering Prize, by Funda\c{c}\~{a}o para a Ci\^{e}ncia e a
Tecnologia under projects PTDC/EEI-TEL/4543/2014 and DL
57/2016/CP1353/CT0001 and by Instituto de Telecomunica\c{c}\~{o}es
under project UID/EEA/50008/2017.
% \end{acknowledgement}

%%%%%%%%%%%%%%%%%%%%%%%%%%%%%%%%%%%%%%%%%%%%%%%%%%%%%%%%%%%%%%%%%%%%%%%%%%%%%%%%%%%%%%%%%%%%%%%%%%%%%%%%%
%%%%%%%%%%%%%%%%%%%%%%%%%%%%%%%%%%%%%%%%%%%%%%%%%%%%%%%%%%%%%%%%%%%%%%%%%%%%%%%%%%%%%%%%%%%%%%%%%%%%%%%%%
\appendix
\section{Wave-equation in a pseudo-Tellegen medium } \label{sec:general_wave_eq_pseudo-Tellegen}
We consider a nonreciprocal material described by the tensors
\cite{serdyukov_electromagnetics_2001}:
\begin{align}
\db{\eps}&=\eps_\parallel \left( \hat{\vec{x}}\hat{\vec{x}} + \hat{\vec{y}}\hat{\vec{y}} \right)+ \eps_{zz} \hat{\vec{z}}\hat{\vec{z}}, \\
\db{\mu}&=\mu_\parallel \left( \hat{\vec{x}}\hat{\vec{x}} + \hat{\vec{y}}\hat{\vec{y}} \right)+ \mu_{zz} \hat{\vec{z}}\hat{\vec{z}}, \\
\db{\xi}&=\db{\zeta} =   \boldsymbol{\xi} \otimes \hat{\vec{z}}
+\hat{\vec{z}} \otimes \boldsymbol{\xi} \label{E:xi_appendix}.
\end{align}
The symmetric magnetoelectric coupling tensors determine a
nonreciprocal pseudo-Tellegen response. Here $\eps_\parallel$,
$\eps_{zz}$, $\mu_\parallel$, $\mu_{zz}$ are the parallel (in plane)
and perpendicular (out of plane) components of the relative
permittivity and permeability respectively, and
$\boldsymbol{\xi}=\xi_x \hat{\vec{x}}+ \xi_y  \hat{\vec{y}}$ is a
vector lying in the $xoy$ plane.

In the absence of current-sources and for a time-harmonic variation
of the form $\e{-i\omega t}$, the Maxwell equations
\eqref{E:Maxwell_eqs} in this system reduce to
\begin{align}
\nabla \times\vec{E} =   i \omega \left( \frac{1}{c} \left( \boldsymbol{\xi} \otimes \hat{\vec{z}} +\hat{\vec{z}} \otimes \boldsymbol{\xi} \right) \cdot \vec{E} + \mu_0\left( \mu_\parallel \left( \hat{\vec{x}}\hat{\vec{x}} + \hat{\vec{y}}\hat{\vec{y}} \right)+ \mu_{zz} \hat{\vec{z}}\hat{\vec{z}} \right)\cdot \vec{H} \right), \label{E:pseudo-Tellegen_eq1} \\
\nabla \times\vec{H} = - i \omega \left( \eps_0 \left(
\eps_\parallel \left( \hat{\vec{x}}\hat{\vec{x}} +
\hat{\vec{y}}\hat{\vec{y}} \right)+ \eps_{zz}
\hat{\vec{z}}\hat{\vec{z}} \right) \cdot \vec{E} + \frac{1}{c}\left(
\boldsymbol{\xi} \otimes \hat{\vec{z}} +\hat{\vec{z}} \otimes
\boldsymbol{\xi} \right) \cdot \vec{H} \right).
\label{E:pseudo-Tellegen_eq2}
\end{align}
Remarkably, for a medium invariant to translations along the
$z$-direction ($\frac{\partial}{\partial z}=0$) the pseudo-Tellegen
magnetoelectric coupling considered here does not mix the
polarizations, i.e, the decomposition into TE ($\vec{E}=E_z
\hat{\vec{z}}$) and TM ($\vec{H}=H_z \hat{\vec{z}}$) waves is valid.
Then \eqref{E:pseudo-Tellegen_eq1} and \eqref{E:pseudo-Tellegen_eq2}
lead to the two following independent equations for TE and TM waves
\begin{align}
\left( \nabla   -  i  \frac{\omega}{c}  \hat{\vec{z}} \times \boldsymbol{\xi}  \right) \times \left[ \frac{1}{\mu_\parallel} \left( \nabla  - i \frac{\omega}{c}    \hat{\vec{z}} \times \boldsymbol{\xi} \right) \times (E_z \hat{\vec{z}})  \right]   &=   \frac{\omega^2}{c^2}  \eps_{zz}  E_z \hat{\vec{z}}, \\
\left( \nabla   + i  \frac{\omega}{c}      \hat{\vec{z}} \times
\boldsymbol{\xi}  \right) \times \left[\frac{1}{\eps_\parallel}
\left( \nabla  + i  \frac{\omega}{c}     \hat{\vec{z}} \times
\boldsymbol{\xi} \right) \times ( H_z \hat{\vec{z}} )\right]
&=\frac{\omega^2}{c^2}\mu_{zz} H_z \hat{\vec{z}}.
\end{align}
The above equations can be further simplified if one also assumes
that the parallel components of the permittivity $\eps_\parallel$ or
permeability $\mu_\parallel$ are space independent, leading to the
following uncoupled wave equations for TE and TM waves respectively
\begin{align}  \label{E:wave_equation_TE}
\left[\left( \nabla   -  i  \frac{\omega}{c}   \hat{\vec{z}} \times
\boldsymbol{\xi}  \right)^2 + \frac{\omega^2}{c^2}
\mu_\parallel\eps_{zz}\right]  E_z    &=     0,
\\  \label{E:wave_equation_TM}
\left[\left( \nabla  + i  \frac{\omega}{c}     \hat{\vec{z}} \times
\boldsymbol{\xi}  \right)^2   +     \frac{\omega^2}{c^2}\mu_{zz}
\eps_\parallel \right] H_z     &=0.
\end{align}
%

% EDITED
% \bibliographystyle{ieeetr}
\bibliographystyle{naturemag}
% \bibliography{Biblio}
\bibliography{../Biblio_CLEAN}

\end{document}